# EVERYDAY CYBER SECURITY IN ORGANISATIONS

## Literature review

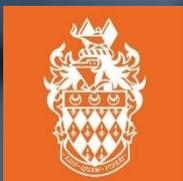

April 2018

Royal Holloway University of London

# Everyday Cyber Security in Organisations

Literature review

## Researchers and principal authors:

Amy Ertan, PhD student, Information Security Group
Georgia Crossland, PhD student, Information Security Group
Claude Heath, Post-Doc, Information Security Group

Professor David Denney, School of Law

Dr Rikke Bjerg Jensen, Information Security Group

Draft review submitted to the Cabinet Office on 24st April 2018.

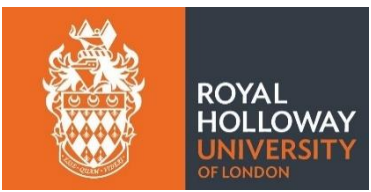

# CONTENTS

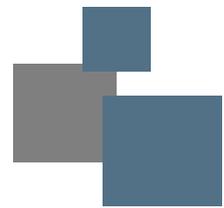

# Recommendations for future research

| | |
|---|---|
| The overarching direction for future research is the need a better understanding of how to foster behaviour change in the context of everyday cyber security. This work should repeat existing research in different environments to determine the validity and reliability of existing methods to a greater extent. | #1 |
| Incoming regulations pose a significant challenge to information security management within organisations. New requirements such as the General Data Protection Regulation (effective from May 2018) have prompted major change programmes across organisations. The effects of the regulation on security behaviours in the workplace have not been specifically examined. | #2 |
| There is a need for more research on behavioural differences between types of employees, or within different organisational environments, as these may also display different behaviours towards cyber security issues. | #3 |
| A common theme within the academic literature was a need to continue to bring together diverse approaches to better understand cyber security behaviours and practices. The theories used in the research of information security draw on a number of distinct fields with evolving theories. | #4 |
| The need for effective security education training was highlighted across the literature. Several academics highlighted opportunities for needs-based analysis to be incorporated into educational policy. | #5 |
| The lack of theoretical underpinnings and critical reflection with the topic of security behaviours in most of the studies, makes the evidence base inconsistent and poor. As a result, it is not possible to recommend any conclusive suggestions as to 'what works'. Greater engagement with the fundamental principles and theoretical framings of security is needed. | #6 |

# Everyday Cyber Security in Organisations

**EXECUTIVE SUMMARY**

Everyday cyber security comprises the security worries that emerge from the micro and proximate interactions, routines, rhythms and actions of everyday life as well as those that link the digital (or cyber) with the individual security concerns. Everyday cyber security therefore concerns both technological security as well as human security. Technological security controls focus on maintaining the integrity of the technology, ensuring usability of technological security, protecting from malware, and controlling access. Human security, on the other hand, is framed within interactions between people mediated through technology, including the sharing of security practices and the negotiation and navigation of convenient and/or effective everyday security practices.

The interrelation between technological and human security in the context of the everyday, within organisational contexts, relies on a series of interwoven developments. First, implementing efficient controls in changing cultural, legal and social landscapes challenges the legitimacy to influence patterns of information production and sharing. Second, technological dependency introduces an array of security challenges as the separation between the externalised security rationale and the internal security dialogue is reduced e.g. if security policies (externalised security rationale) are seen to be unfair or 'too time-consuming' people are likely to find more convenient ways of using technology (internal security dialogue). Third, the pressure of always being 'on' highlights the importance of understanding everyday security practices and the challenges imposed by the integration of technology in most aspects of life and work.

**Introduction to the review**

This review explores the academic and policy literature in the context of everyday cyber security in organisations. In so doing, it identifies four behavioural sets that influences how people *practice* cyber security. These are: compliance with security policy; intergroup coordination and communication; phishing/email behaviour; and password behaviour. However, it is important to note that these are not exhaustive and they do not exist in isolation. In addition, the review explores the notion of security culture as an overarching theme that overlaps and frames the four behavioural sets. The aim of this review is therefore to provide a summary of the existing literature in the area of 'everyday cyber security' within the social sciences, with a particular focus on organisational contexts. In doing so, it develops a series of suggestions for future research directions based on existing gaps in the literature. The review also includes a theoretical lens that will aid the understanding of existing studies and wider literatures. Where possible, the review makes recommendations for organisations in relation to everyday cyber security.

To this end, the review aims to:

- Examine existing social science literatures on everyday cyber security behaviours within organisations.

- Identify key behavioural sets that influence how individuals engage with and use technology in the workplace; including the individual behaviours that relate to the behavioural sets.

- Outline the motivating factors that drive 'good' behaviour in relation to the key behavioural sets.

- Identify real and perceived barriers to behavioural change in the context of everyday cyber security as understood in relation to the specified behavioural sets.

- Set out recommendations for future research approaches.

- Highlight approaches which have worked in previous studies focusing on behavioural change in organisations.

# SECTION ONE:
# INTRODUCTION TO THE LITERATURE

Cyber security continues to be a central concern within organisations, with cyber-criminal activity posing a significant risk and costing vast amounts of money. This is, however, not surprising given the number of security-related threats facing organisations as well as employees on a daily basis (Sommestad et al., 2014). Recent reports suggest that cyber-attacks, whilst expensive for businesses also shake the public's and the consumer's confidence in the ability of organisations and governments to keep information safe (Nandi et al., 2016).

Some cyber-attacks will be unknowingly enabled by employees, for example by falling victims to phishing attacks (Krombholz et al., 2015). Government reports suggest that in 2015, 90% of all companies suffered a security breach, with 75% of large businesses suffering a staff-related security breach (HM Government and PWC, 2015). Moreover, of the worst breaches in the same year, 50% were found to have been caused by inadvertent human error (HM Government and PWC, 2015). Although some of these employee behaviours may be mitigated by technological advances and interventions, humans interact with technology as part of their everyday (Martins et al., 2014) and so technological solutions are, at least in this way, limited. Human elements have also been of importance in other disciplines with a human-technology interaction aspect, such as aviation, where there is a larger pool of knowledge about potential risks posed by human errors (e.g. Helmreich and Foushee, 1993). This goes beyond wider literatures on Human-Computer Interaction (HCI) as it has security at the centre.

Most organisations will develop cyber security policies in an attempt to both communicate envisioned threats and risks to employees as well as set standards for how employees should behave in a cyber environment. However, it is critical that organisations understand the human-behavioural factors within cyber security management processes (Parkin, van Moorsel and Coles, 2009). There has been a growing interest in investigating the issue of employee behaviour and attitudes and what implications certain behaviours and attitudes may have for security within organisations (Guo, 2013). In 2014, for example, a systematic review of quantitative research analysed 29 studies to identify variables that influence compliance with information security policies (Sommestad et al., 2014). The results showed that over 60 variables, many of which are incorporated in this review, have been studied in relation to information and cyber security policies. This, if nothing else, demonstrates the diversity of influences on human behaviour in a cyber security context. Moreover, it emphasises the focus on quantitative approaches to exploring cyber security behavioural traits, which has dominated much research in this area.

Still, literature on this topic draws from many different disciplines, mainly those of a social science nature. These disciplines include, but are not limited to psychology (West, 2008), sociology, criminology, and economics (Theoharidou and Gritazalis, 2007). Within the psychological literature, the influence of personality types on cyber security behaviours has been explored in different settings (e.g. Whitty et al., 2015). This said, generally it has been found that personality plays a small role in how people behave online, and companies cannot change their employees' individual personalities. Therefore, personality will not be the focus of this review. Rather, the focus will be on the key behavioural sets that have emerged in the wide range of social science literatures that have been reviewed in response to the overarching aims of this study.

**Structure of the Review**

The main body of this literature review first explores the notion of security culture as an overarching framework for understanding everyday cyber security practices, before addressing four key behavioural sets that have arisen from reviewing existing studies and theoretical frameworks. The summary sections outline 'what works' in terms of changing security behaviours as well as setting out the limitations of the literature and outlining future research directions, based on existing literatures and studies.

## SECTION TWO:
## SECURITY CULTURE

Before exploring the individual behavioural sets identified in the existing body of literature, this section discusses how the notion of security culture has been approached and understood in such literatures focusing on cyber security behaviours.

**Definitions and context based on the literature**

Like culture, security culture is a contested concept. There is lack of consensus in relation to the composition and definition of security culture within the literature. Security cultures are influenced by a range of factors within and beyond organisations, including positive and negative security behaviours and practices. Positive security is defined in the literature as the freedom to go about daily life, where security has been successfully negotiated by groups and individuals (McSweeney, 1999). It compares with what might be called 'negative security' which refers to freedom from threat (protection). McSweeney (1999, pp.94) notes that "situations of security breakdown cannot be considered the only litmus-test of our conception of security", as human needs go beyond this and also concern agency and moral choice. This approach favours "cooperation, inclusiveness, and the positive amelioration of intergroup relations" as potentially useful ways to ameliorate conflicts of interest and practice (McSweeney, 1999, p.98).

In a broad sense, every organisation has a particular culture, consisting of an omnipresent set of assumptions that directs the activities within the organisation, for example, those directed towards security (Van Niekerk and Von Solms, 2010). Such a security culture may refer to a set of norms and values developed and shared by members of the organisation towards different aspects of security (D'Arcy and Greene, 2014), which determines the mind-set of members towards security within the organisation. However, norms and values are difficult to quantify and researchers must be careful not to oversimplify the term 'security culture'. For example, it is important to note that members need to identify with the culture in order for it to foster the desired security mind-set.

There are many different academic papers and frameworks describing what is meant by a 'positive' cyber security culture (e.g. Martin et al., 2006; Ruighaver et al., 2007; Schlienger and Teufel, 2003). Taking such papers into account, an effective security culture would ideally involve adhering to security policies, reporting when things appear suspicious or go wrong and feeling comfortable to do so, as well as making security a priority across all levels of an organisation. Considered more broadly, a positive or 'good' workplace culture is understood to increase employee commitment and loyalty to the organisation (e.g. Martin et al., 2006), which may in turn foster a better security culture. In contrast, a negative or 'bad' security culture may encompass a lack of understanding of security, a blasé attitude towards security, and 'non-compliance' towards security measures.

Security culture is therefore important to both understand and encourage in organisations because if it is positive, employees are more likely to see the importance of cyber security controls and practices, believe in and be committed to enforcing them (Parsons et al., 2015; Parsons et al., 2010; Renaud and Goucher, 2012).

However, security culture cannot be looked at on a singular level as there are also subcultures within organisations (Kolkowska, 2011), and such subcultures may transcribe to different and conflicting values. Research highlights that value conflicts are important factors to take into account when security culture is developed in an organisation (Kolkowska, 2011). This links to the third behavioural set outlined in this literature review; namely, intergroup coordination and communication. Of course, there are other factors that may contribute towards whether someone adheres to controls. These include a country's cultural norms and trust in the idea that the organisation protects the employees (Hovav and D'Arcy, 2012). Security culture does therefore not exist in isolation but builds on wider cultural and societal traits.

**Security culture and 'good' cyber security practice**

Owing to its definition, security culture is generally looked at in the literature as a broad behavioural term, used to describe workplace attitudes, reactions, activities and mindsets. It therefore also encompasses compliance with security policy, password behaviour and phishing/email behaviour, and influences intergroup coordination and communication, which are the four behavioural sets that form the main part of this review.

Researchers argue that security culture is an important factor in maintaining an adequate level of security in organisations, and contend that only a significant change in security culture can tackle the broader 'human aspect' element in security breaches (Hovav & D'Arcy, 2012). This may involve attempting to change employee past automatic (habitual) behaviours, changing cognitive heuristics, and developing psychological contracts that encompass security. To this end, if a security culture is poor, it may foster poor security behaviours in a broader sense. These poor security behaviours may include those in this review, such as bad password behaviour.

In the context of some organisations, researchers have argued that there may be a culture of too much trust in the security systems. To this end, employees may believe that the IT systems in place protect them from any cyber-attack, and in doing so may feel less responsible for computer related security issues (Benson et al., 2018). This trust in security systems therefore may lead to misguided security practices such as trust in email providers to catch phishing emails.

Habitual and past automatic behaviours relate to security culture as such behaviours make up, and are often the consequence of, a particular cultural attitude. Therefore, changing security culture in an organisation may include attempting to change the individual habits of employees. Vance et al. (2012), in a study looking at the effects of habits on compliance with information security policies, found 'habit' to be an important role in the context of employees' compliance with information security policies. Habit was also found to have a significant influence on whether employees felt they were subjected to threat if they did not comply with information security policies. Habitual behaviour in IT use does not always require conscious behavioural intention (de Guinea and Markus, 2009), and so employees may perform unsafe cyber security behaviours without an explicit awareness of the potential consequences of their actions.

**Cultural change**

Owing to the fact that security culture encompasses many aspects of organisational behaviour, such as both knowledge and attitudes towards information security policies or procedures, changing security culture may prove a difficult task (Harris and Ogbonna, 1998). However, the literature does highlight some possible ways for cultural change.

Firstly, research shows that a positive security culture can benefit from being led from the very top of an organisation – requiring executive level colleagues to show commitment to and take ownership of security-related issues (Hovav & D'Arcy, 2012). Similarly, this top-down approach needs to be understood and accepted throughout the organisation. It is therefore important that security policies and guidelines are accompanied with clear narratives and messages. A study by Hu et al. (2012) found that top management participation in information security initiatives have significant influence on employees' attitudes towards compliance with information security policies, and strongly influences organisational security culture. This demonstrates how top management can play a proactive role in shaping culture within organisations. Additionally, top-management participation in security practices and training will also improve the quality of management of information security according to the literature (Soomro et al., 2016).

The examined literature also builds on a lot of research on awareness campaigns. Awareness campaigns are based on the premise that employees should be made aware of their security responsibilities, and take them seriously, if they are to be expected to comply with them. The literature suggests that this can be achieved by making it clear that their actions (or inactions) can have serious consequences to the organisation as well as themselves. This is generally implemented in organisations through security awareness programmes (Guynes and Windsor, 2012). Increasing awareness is seen to increase employee responsibility and highlight likely security threats to staff. (Von Solms and Von Solms, 2004). However, if we look at Protection Motivation Theory (Prentice-Dunn and Rogers, 1997, as cited in Floyd et al., 2000), and the research surrounding fear-appeals, it becomes clear that aggressive 'scare tactics' in awareness raising initiatives may have the opposite effect than the one intended, leading to counter-behaviours ultimately damaging the security culture of an organisation. Furthermore, although awareness initiatives have been used and researched globally (Chen et al., 2008), security awareness plans are limited to organisations, or sectors of organisations, that do not have an existing awareness of cyber security. Many companies may indeed already have high awareness of the company's issues and responsibilities surrounding cyber security.

Research in this area has also looked into the use of incentives, such as sanctions and rewards, to change behaviour and shape culture (Herath and Rao, 2009). Such attempts of changing behaviour will be looked at in more detail within some of the four behavioural sets set out in the next sections, as research in this area tends to focus on changing specific behaviours or changing compliance levels rather than changing culture more broadly. Research in this area have reasonably mixed findings, with the outcomes of rewards and sanctions depending to a certain extent on other environmental factors, and the degree to which employees are rewarded or sanctioned. Some research in this area has suggested that the use of sanctions and rewards on employees have little effect on employee behaviour towards information security (Pahnila et al., 2007). This may be due to a variety of reasons (Glaspie and Karwowski, 2017). For example, it has been suggested that rewards often fail to work if the benefit of non-compliance outweighs the perceived incentive (Vance et al., 2012). Furthermore, Farahmand et al. (2013) argue that incentives should only be used if they influence a large number of people to act for the common cause, and other incentives are inefficient. Furthermore, other researchers have argued that if prosocial behaviour is developed within an organisation, the need for sanctions or rewards to influence compliance with security policy is eliminated (Thomson and van Niekerk, 2012). In a prosocial workplace, employees are concerned about and adherent to information security policies.

In relation to this, researchers have also speculated the extent to which the new General Data Protection Regulation (GDPR) will have an effect on the way companies approach and deal with cyber security (Becker et al., 2017). Psychologists and other disciplines may look at the effect of relevant regulations and make predictions about whether financial penalties will assist in encouraging the development of a better employee security culture. This will likely be affected by the extent to which employees feel the GDPR is in their interest (Fritsch, 2015).

The use of 'security champions' has also been investigated in the literature. Security champions are a way to promote and monitor security policy at an employee level, encouraging local representatives to demonstrate and lead by example (Becker et al., 2017). Gabriel and Furnell (2011) suggest a security champion should promote and foster awareness, motivation, and compliance. The concepts driving security champions stems from the idea of the influence of social control on employee's security habits (Hsu et al., 2015). Research shows that formal control and social controls, both individually and collectively, can enhance both role specific and extra-role security behaviours (Hsu et al., 2015). Research has also shown that security champions need to take into account the differing viewpoints and subcultures across organisations, and not miss the opportunity to engage the wider organisation (Becker et al., 2017). In this way, security champions can be used as 'bottom-up' agents, to change better improve policy, rather than working from a 'top-down' approach to ensure compliance with existing, and possibly flawed policy (Becker et al., 2017).

The theory of 'behavioural nudging' (Thaler & Sunstein, 2008) has also been studied in the cyber security context. The theory suggests that people can be nudged towards certain choices and behaviours, given certain environmental cues, without forcing outcomes on anyone (Benson et al., 2018; Coventry et al., 2014). Therefore, applying this theory to cyber security in the workplace may be useful in leading to more vigilant behaviours by employees. This theory has also been applied to try and nudge the public towards using more secure wireless networks (Turland et al., 2015).

**Limitations and Conclusions**

It is becoming widely accepted in the literature that creating a positive information security culture is key to maintaining healthy security behaviours in the workplace (Karyda, 2017). However, there are some major gaps in the research within this area. Firstly, research focusing on defining and measuring the cybersecurity culture is lacking (Gcaza and von Solms, 2017). There is also a lack of valid data and studies on what initiatives actually work to change culture. Within the existing body of knowledge, the review has highlighted a number of additional research gaps and needs as also highlighted by Karyda (2017):

- there is a need to explore subcultures further;
- the need for uncontested definitions of information security culture;
- the need to look at the crossovers and effects between organisational structure and management practices and their effects on security culture;
- the need to research the impact of overall organisational culture, comparisons between security culture raising programs;
- and the effects of employee culture outside the workplace on security culture.

This list is not exhaustive and highlights the plethora of open questions and opportunities for future research on security culture and related themes. At a time when most organisations are referring to the need to change security culture and security mindsets, it is particularly important for researchers to begin to fill existing knowledge gaps.

## SECTION THREE:
## BEHAVIOURAL SETS

This section sets out the four behavioural sets that have been identified in the literature as the dominant behaviours. Whilst these sets are not exhaustive, they also do not exist in isolation. This means that there are a number of overlaps between the different behavioural sets and that they include aspects of other behavioural sets not covered in this review.

Each behavioural set is divided into five sub-sections which (1) introduces key definitions and context based on the literature, (2) sets out specific behaviours related to the overarching behavioural set, (3) identifies motivating factors driving 'good' behaviour in the context of the specific behavioural set, (4) outlines identified barriers to behavioural change, and (5) recommends 'what works' as gleaned from the literature.

### 'Compliance' with security policy

The first behavioural set identified in the literature relates to compliance in the context of security policy within organisations.

*Definitions and context based on the literature*

Compliance refers to 'the state or fact of according with or meeting rules or standards' (Oxford Dictionary). Compliance in relation to security policy and procedures is a hot topic in most organisations. This particularly applies at the board level as organisations face an increasing amount of statutory and regulatory requirements, such as the GDPR. While measuring overall compliance levels is a complex task, under GDPR executives will face renewed and intense pressure to achieve high levels of employee compliance as part of wider security goals.

In the context of security, it is widely accepted that compliance with adequate information security policies will lead to a more secure information security level within an organisation. However, achieving an ideal security policy compliance level is a complicated task (Sommestad et al., 2014), and a range of literature highlights that even when information security policies and documents are in place within an organisation, its employees do not necessarily comply with its requirements (Ifinedo, 2014; Hazari et al., 2009). In some cases, policies are viewed as merely guidelines rather than mandated instructions, while employees may also choose not to comply for reasons of convenience when carrying out day-to-day roles (Herath & Rao, 2009).

In fact, over half of all information security breaches are estimated to be indirectly or directly caused by employee failure to act in accordance with information security procedures (Stanton et al., 2005). In addition to academic research, frequent security incidents and industry surveys highlight the difficulties of enforcing security through compliance, and suggest that while policies and procedures may be in place, many employees and outside contractors will not comply. Additionally, Federal Bureau of Investigation (FBI) survey reports have highlighted end-user policy as one of the key challenges in achieving target levels of information security (Herath and Rao, 2009).

As a complex topic, understanding compliance within an information security policy environment should be a cross-discipline exercise, including but not limited to research from the psychology and criminology fields (Aronson et al., 2010). As explained below, a dominant approach stems from the Theory of Planned Behaviour (TBP), encompassing Social Cognitive Theory (SCT) as a framework to explain behaviours including information security policy compliance (Hazari et al., 2009). Examining mechanisms of behavioural change through theories such as these, researchers have identified factors that have significant impact on employee beliefs and attitudes to information security policy, as well as employee intention to comply.

As well as explaining employee behaviours and motivations for poor information security compliance, a number of a behavioural approaches explore methods for improving employees' compliance with the security procedures of their organisation. Campaigns to inform employees on threat severities, as well as campaigns addressing employee self-efficacy, social norms within the organisation and wider security culture, and comprehensive enforcement systems all propose to impact compliance behaviour.

*Behaviours related to the behavioural set*
Employees fail to comply with information security procedure for a number or reasons. Several socio-cognitive theories explain why an employee may, implicitly or deliberately, violate security principles, and many of the targeted behavioural approaches, are interdisciplinary in nature. Research draws on a combination of rational choice and decision-making theories, deterrence theory, and other principles from criminology, psychology and other research areas to explain behaviours and attitudes to compliance. In addition, an examination of an employees' perceived compliance burden as well as detrimental impact on productivity lead to the emergence of 'shadow security' behaviours (Kirlappos et al., 2014).

Employee attitudes towards compliance have been shown to determine intention to comply with security policy. Applying rational choice theory (the principle that individuals tend to make logical decisions), an employee's attitude is influenced by benefits of compliance, the cost of compliance, and the cost of non-compliance. The employees' beliefs significantly affect employees' assessment of consequences, which in turn affect an employee's attitude. (Bulgurcu et al., 2010). Perceptions on the actual and anticipated costs and benefits of compliance to the employee and to the organisation are key factors in the compliance decision. The individual will consider their own needs rather than the more 'altruistic' process of compliance (Beatement et al., 2008).

Habit and experience have been shown to strongly affect decision-making when it comes to employee compliance, explained through Protection Motivation Theory (PMT), which explains how individuals determine a reaction to a threat, or 'fear appeal' (Vance et al., 2012). The importance of past behaviour and habit are highlighted through PMT and involve a cognitive process to perceive threats, looking at the potential rewards of the threat, its severity (magnitude of the threat), and vulnerability (susceptibility to the threat). It also includes cognitive processing of factors enabling an individual to deal with a threat: response efficacy (perceived benefits by acting to remove the threat), response cost (to the individual for implementing the protective behaviour) and self-efficacy (the belief that the protective behaviour may be implemented). In the context of information security within an organisation, Vance et al. (2012) research found nearly all components of PMT significantly impacted employee intention to comply with information security.

Deterrence on non-compliant behaviour relates to organisational punishment mechanisms and sanctions in this context. Deterrence theory suggests that certainty, severity and celerity (swiftness) of penalties affect people's decisions on whether to commit a crime or not. In a seminal study applying deterrence to information security, classical deterrence theory suggested stating penalties for information security policy non-compliance increases security behaviour (Straub, 1990), however, this assertion has been challenged by other studies (e.g. Pahnila et al., 2007).

Finally, the idea of compliance as a binary decision ('complying, or not') has been challenged by Kirlappo et al. (2014) who highlighted a third response: "shadow security". Shadow security practices emerge where security-conscious employees believe they cannot comply with a prescribed security policy, and therefore create a more fitting alternative to the policies suggested by the organisation's official security functions. These workarounds are usually not visible to official security employees or higher management. This behavioural pattern emerges from the conflict employees face in trying to carry out their roles while managing risks as far as possible, representing a compromise that may or may not be as secure as the official policy. It represents a response where security policy has become problematic or overly onerous on an employee.

*Motivating factors driving 'good' behaviour*
Social influences have been shown to be greatly influential in forming employees' intention to comply with security procedure, and can be explained through the Theory of Planned

Behaviour (TBP) approach. The TPB examines social influence, referring specifically to a change in an individual's attitudes or behaviours as a result of interactions from another individual or group. The theory proposes that individual behaviour is influenced by attitudes (positive or negative feelings towards engaging in a specified behaviour), subjective norms (an individual's view of what people important to them would think about a given behaviour) and perceived behavioural control (the individual's beliefs on the resources needed to facilitate a behaviour) (Ifinedo, 2014). Top management participation in information security initiatives has been shown to have direct and indirect influence on employee attitudes towards information security policies

Social norms also apply through social bonding theories relating to social control. Social control is a sociological concept referring to a mechanism that regulates individual and group behaviour, leading to compliance with the rules of a given group. In a control context, employees' intentions have been shown to be significantly influenced by co-worker behaviours, as well as social pressures exerted by subjective norms (Cheng et al., 2013). These suggest that the influence of informal social control, based on customs and social values and implemented by unofficial controlling individuals, can have a real impact on policy compliance. As organisations are also social groups, the research carried out by Cheng et al. (2013) suggests that the same social control mechanisms are applicable to organisations. The effects of attachment to one's organisation and job are significant, while the expectations of colleagues, managers and significant others showed significant influence on employees' policy violation intentions (Cheng et al., 2013).

Compliance is also likely to be affected by an employee's attitudes to their employer overall, with loyalty being an important theme. More specifically, the 'psychological contract' between the employee and the employer, which refers to any unwritten expectations and to how the relationship between the employee and their employer is perceived, has shown that feelings of loyalty and willing compliance may also influence security policy compliance (Han et al., 2017). To this end, the study of information security compliance necessarily requires a thorough understanding of employee and employer motivations and the assumptions that go along with these, in order to fully understand decisions about which practices are followed.

Factors such as employee threat perception on the severity of breaches along with response perceptions or response efficacy, self-efficacy and response costs all affect attitudes to security policy, and have an impact on policy intentions. Information quality and security awareness have been shown to impact on employee compliance. Information security awareness positively affects both attitudes and an employee's outcome beliefs, leading to an increased intention to comply with information security policy (Burgurcu et al., 2010). Organisational commitment is able to impact intentions through promoting believes that employees actions have an effect on the organisation's overall information security. (Herath and Rao, 2009).

*Barriers to behavioural change*
The key barriers to greater compliant behaviour levels are poor security culture and employees' negative attitudes to compliance, which may be further entrenched by a number of factors, including unmanageable sanctions and reward enforcement of sanctions and reward, as well as the perceived hindrance of compliance. Results from survey data collections within organisations provide empirical support that security culture is a driver for employee security compliance within an organisation. (D'Arcy and Green, 2014). If employees perceive the organisation to have a lax security culture and not to prioritise information security policy compliance, the data suggests that it will follow that compliance will be lower than organisations with a more proactive culture.

Research output on disciplinary methods and punishment suggests mixed findings on the deterrence of poor compliance. Research focus on compliance in general suggest sanctions and penalties as a deterrence mechanism, implicitly suggesting that when security policy violations are severely punished, employees will no longer carry out the offending actions. However, there are a number of counter-arguments suggesting that this is not necessarily effective in theory or in practice. Kirlappos et al. (2014) outline that the challenges of complete monitoring across widespread activities and disciplining a large number of employees, who are known to breach compliance policy, make it an unmanageable task to carry out effectively. Without a very significant investment of financial and human resources, sanctions cannot be enforced evenly, and sanctions that are not enforced are not an effective deterrent and do not impact compliance rates. Furthermore, heavy-handed enforcement can prove a barrier to effective behavioural change, instead promoting tensions between security enforcers and the rest of the organisation (Kirlappos et al., 2014). Forcing conformance to compliance policy through disciplinary methods has also been shown to lead to inefficiency in terms of 'compliance delay'. As well as negative attitudes to a policy, employees may wait until the last possible opportunity to comply, potentially causing operational issues. (Belanger et al, 2017). Examples of this may include users accidentally becoming locked-out of a system or application having forgotten to reset their password in the mandated window, or a potential application delay if employees try to complete mandatory online training en-masse.

On the other hand, persuasion will also be limited if compliance conflicts with employee productivity in their roles. If employees feel that complying with the security policy will prove to be a hindrance to their day-to-day activity, with burdensome mechanisms more likely to slow them down, this will be viewed as 'time wasted by security' and will lead to conscious violation of policy. Frustration on behalf of employees leads negative attitudes to the policy which will influence compliance intention. (Herath and Rao, 2009; Kirlappos et al., 2014). These attitudes subsequently limit the effectiveness of awareness and educational campaigns as employees discredit information and may even progress to discourage compliance even with policies that cause minimum convenient as 'it all adds up' (Kirlappos et al., 2014).

*Recommendations and 'what works'*

Parsons et al.'s (2015) research highlighted a small to moderate positive relationship with information security culture and employees' information security decision-making. Encouraging improvements to an organisation's security culture was suggested to benefit security management programmes and the behaviour of employees, which in turn should improve compliance with security policies (Parsons et al., 2015). Several aspects within a security culture are suggested to impact the intentions and motivations of employees; a complex set of interactions that may refer to security communications, effective monitoring and sanctions enforcement and management commitment. Education and awareness campaigns should take into account employees past and automatic behaviour in attempts to achieve behavioural change (Vance et al., 2012), while deterrence and social bonding theories suggest that subjective norms and co-worker behaviours significantly influence employee intentions. Developing and launching initiatives to encourage positive attitudes towards security policy and establish consensus-building programmes can utilise protection motivation theory and social control theories to encourage compliance.

Overall, research is mixed when it comes to sanctions and rewards, with no conclusive evidence highlighting intrinsic virtues in either. Findings suggest that neither certainty of sanctions (how likely the sanction is to be enforced) nor application of rewards appear to have any significant effect on compliance (Pahnila et al., 2007; Cheng et al, 2013). While organisations have a far stronger emphasis on penalties, with indications that high-cost penalties resulted in improved employee knowledge of policies, this did not translate to any increase in attitude or instances of self-reported behaviour (Parsons et al, 2015). The perceived severity of sanctions was seen to impact intention to comply, suggesting organisations would benefit from clear declarations on enforcement structures and punishments for policy violations (Cheng et al., 2013). Reward enforcement may be an alternative for organisations where sanctions do not successfully prevent violation, supporting a wider general deterrence theory. There are significant interactions within punishment and reward and (deterrence theory more broadly) that indicate a need for a comprehensive enforcement system, through which a reward enforcement scheme can help establish and emphasize organisation moral values (Chen et al., 2012).

The employee relationship is another key area where improving dynamics can lead to positive behavioural change. This is supported by further studies which highlight that employees' attitude, normative beliefs and habits all influencing the intention of employees to comply with security policy. While these can be influenced by a number of factors, there is empirical support to suggest that an employee's feeling of job satisfaction will influence their security compliance intention. These security related general work environment factors contribute to security compliance intention, with higher job satisfaction rations having an increased tendency towards compliant security behaviour (D'Arcy and Green, 2014). The research suggests that organisations that create a work environment where employees are satisfied will not only benefit through improved quality of life, but increase. Recommendations specifically refer to human resource initiatives such as job enrichment programmes, contributing to job security as a way to drive positive security behaviour (D'Arcy and Green, 2014).

Similarly, social norms and the 'social contract' between employees and employers from the perspective of security theory can illustrate how perceived self-efficacy may impact compliance. Practitioners should utilise the influence of positive social pressure (normative beliefs) from executives, supervisors and peers to form a social environment that best encourages positive security behaviours (Pahnilo et al., 2007; Hu et al, 2012). In terms of the theory of the social contract, this is generally understood to be accompanied by the freedom of the individual to order actions and dispose of resources as thought fit to ensure security. Resource availability is a factor that can significantly affect the perceived self-efficacy of an employee when it comes to security, which in turn predicts compliance policy intentions. This has implications for how shadow security practices may be approached as having a potentially positive impact on security if negotiated correctly. It is suggested that organisations should try to learn from shadow security behaviours, assisting the development of a 'workable' security policy that offers security without impeding on the organisation's business (Kirlappos et al., 2014).

**Intergroup coordination and communication**
This second behavioural set identified in the literature relates to intergroup coordination and communication within organisations.

*Definitions and context based on the literature*
A group refers to a collection of people with shared characteristics (Brown, 1988). Such as those within a peer-group, school, or organisation. The study of inter-group coordination and communication looks at interaction between two or more collections of groups. Disagreements between groups may strengthen a group divide and miscommunication (Brown, 1988). Therefore, dynamics and coordination within and between work teams can have a significant effect on the ways in which an organisation *practices* and communicates cyber security.

Intergroup coordination and communication have been studied in a wide array of disciplines with a variety of methodologies. Economics, sociology (both Functionalist and Marxist), psychology and psychiatry are all represented in the diverse research on this topic (Nelson, 1989). A sector of this research has looked into the importance of intergroup relations within organisations, and their effects on organisational functioning. Group norms inevitably develop in organisations, perhaps depending on job level, or department sector. Therefore, conflict can sometimes develop when the groups within the workplace do not communicate effectively.

*Behaviours related to the behavioural set*

In this instance, intergroup coordination and communication may refer to behaviours that develop from the divide between high-level managers and low-level employees – both not understanding how one's behaviours may influence, or are seen by, the other (Albrechtsen and Hovden, 2009). For example, high-level managers may not understand how some policies have practicality issues, or give employees (a feeling of) information overload. One relevant study found that the most frequently used cyber security measures in companies in Norway are technological measures, instead of more effective awareness campaigns (Merete Hagen et al., 2008). If we understand this from an inter-group dynamics perspective, as defined above, this may highlight misunderstanding on the part of management, when it comes to determining what works for employees. Moreover, it suggests that security policies may be based on perceived or imagined behaviours amongst lower-level staff – rather than actual or real behaviours and practices.

Inter-group coordination and ocmmunication also refers to a sort of 'digital divide' (Albrechtsen and Hovden, 2009) between those highly knowledgeable in information/cyber security and employees with less expertise or who perceive themselves to be less knowledgeable. In addition, research has demonstrated that cyber security professionals in some organisations can tend to regard some employees as a potential threat, whereas employees believe that they should be seen and treated as a resource for security practitioners (Adams and Blandford, 2005).

*Motivating factors driving 'good' behaviour*

As demonstrated above, research highlights that a large problem in this area is that security managers and other, perhaps less digitally minded, have different points of view in regard to information security practices (Albrechtsen and Hovden, 2009). These misunderstandings are likely then to be further polarised if the differing groups have no contact or communication with each other. Differing skills, such as those with technological skills and those without, have also been shown by research to polarise groups further (Grugulis and Vincent, 2009). Therefore, rather than anything driving good behaviour here, it seems clear that developing better communication between these groups, and developing understanding about the issues that have arisen, is the key to reducing conflict and addressing different group dynamics.

*Barriers to behavioural change*

It is widely accepted that disagreements between groups may lead to inter-group tensions and may foster poor work relations. As highlighted in the previous sections, such disagreements may arise out of a digital divide between certain employees, or miscommunication and misunderstanding between different levels of staff (Adams and Blandford, 2005). The user of information security systems is often thought of as the enemy by information security staff, however employees feel they are not. The idea that the 'user is not the enemy' (Adams and Sasse, 1999) is a response to the approach to security that highlights users as the enemy, producing clashes between security concerns and users' work efficiency and practices. These differing perceptions therefore need to be overcome, if intergroup relations are to improve in this area.

Multi-national organisations also pose a difficulty to behavioural change in regard to intergroup relations. In certain companies, an information security department may be in another country, and so trying to improve intergroup communication and relations may be difficult. Furthermore, it makes it hard for less digitally minded employees to explain any user problems they might be experiencing. Furthermore, security policies in such organisations, may reflect personal preferences and experiences of those creating them (Siponen and Willison, 2009) - and these preferences may differ between multinational sites. Research also shows that there is a level of disagreement generally between businesses in different countries regarding the top information security problems they face (e.g. Watson and Brancheau, 1991). Additionally, having a multinational company, or even a large multi-departmental company, makes it challenging to develop a baseline level of communication and awareness between groups. Research shows this to be an issue for information security personnel, as effective interactions and communications are necessary for mutual understandings of security risks among different stakeholders and departments (Werlinger et al., 2009).

In addition to the previously mentioned barriers, there may be a level of intergroup prejudice/bias about and between departments and between levels of employees. Intergroup bias refers to the idea that there is a tendency for people to regard members of one's own group more favourably than members of other groups. This can further lead to poor communication between groups and an unwillingness to understand differing views and perceptions.

*Recommendations and 'what works'*

In this area, one argument re-surfaces across the different disciplines looking at intergroup contact - 'The Contact Hypothesis'. In simple terms, the contact hypothesis states that direct contact between groups helps to alleviate conflict (Nelson, 1989). This hypothesis has much supporting evidence in a variety of fields, and research into this has been going on for many years (Allport, 1954; Pettigrew, 1998). There are a few different ways in which researchers believe intergroup contact may reduce conflict. Firstly, if groups interact, they are able to develop positive feelings towards one another. Secondly, group contact reduces a sense of group boundaries and polarisation. Thirdly, contact provides and outlet of disagreement (Nelson, 1989). However, although there is scholarly agreement that contact reduces conflict, it is also noted that the contact should be positive and not competitive. Furthermore, there is not much clarity on how much contact is required to foster positive intergroup coordination and communication; and of course this changes depending on circumstance.

In relation to everyday cyber security, research has shown that managers in companies view the most efficient way of working together and of influencing employee behaviour and awareness to be interaction in some form between users and security managers, e.g. in small face-to-face information meetings (Albrechtsen and Hovden, 2009). However, this method has also been demonstrated to be one of the least frequently used (Albrechtsen and Hovden, 2009). This idea of face-to-face contact between groups (in a physical, rather than virtual sense) to improve relations is therefore related to the 'Contact Hypothesis' (Pettigrew, 1998). Under certain conditions, the contact may act as a positive way to improve relations and communication between groups in organisations regarding cyber security.

The idea of superordinate goals has also long since been studied in the literature as a way to reduce intergroup conflict (Sherif, 1958). Although, from what we have seen in the literature, the idea has yet to be applied to cyber security in the workplace specifically. The idea of superordinate goals is that a set of goals are developed, where it is made clear that to achieve these goals, participation from both/all groups is required. The idea postulates that then, if the goals are achieved, the relationship between the two groups becomes more harmonious (Gaertner et al., 2000). In cyber security, this could be done through highlighting certain goals, like improved password behaviour from employees, includes the whole company, and can only be achieved through information security experts, and general employees, working together. Of course, at this point this is just an idea, and will need to be corroborated by further research.

Leadership, especially effective leadership across work groups, departments and whole organisations, is important in developing relationships between employees. Intergroup leadership (Hogg et al., 2012), also relates to the idea of superordinate goals, as it refers to the idea of leadership across organisational group boundaries. This is of importance because leadership is often required over different formal groups, rather than just one group. Hogg et al. (2012), in their paper on intergroup leadership, argue that developing intergroup identity is key and they identify ways in which leaders can do this. They argue that the leader should champion and be positive about group collaboration and l should also perhaps consider a coalition of leaders (to reduce outgroup feelings by certain groups).

It should also be mentioned in this section that while different groups or company aims may suffer because on intergroup conflict, others may thrive (Bradley et al., 2015). There is research suggesting that under certain conditions, disagreements between teams may improve performance. However, from the literature we gather that this has also yet to be looked at in improving cyber-security practices.

**Phishing/email behaviour**

This third behavioural set identified in the literature relates to phishing and email behaviour within organisations. With most communication within and between organisations and employees taking place over email, and with the increase in phishing attacks and organisations investing significant amounts of money in counter-measures, this is a growing area of research.

*Definitions and context based on the literature*

According to the Anti-Phishing Working Group (APWG): "Phishing is a criminal mechanism employing both social engineering and technical subterfuge to steal consumers' personal identity data and financial account credentials." Social engineering refers to the deception where an attacker attempts to deceive a victim into performing a certain action that benefits the attacker, for example clicking on a malicious link within an email (Mitnick et al., 2002). A method of online identity theft, a phishing attack will, for instance, involve emails directing victims to visit fake replicas of legitimate websites. While automated anti-phishing tools have been developed (for example, Calling ID Toolbar, Firefox 2, ebay Toolbar), they are not entirely reliable (Kirlappos and Sasse, 2012). While security experts and application developers continue to improve phishing and spam detection tools, some authors continue to portray the 'human' as 'the weakest link' in this context (Arachchilage and Love, 2014). As attackers continue to adapt their techniques to social engineer victims to follow a phishing email instruction, researchers have explored user education and employee email behaviour as a means of preventing phishing (Sheng et al., 2007).

Email behaviours are framed within corporate email policy which outlines specific guidelines for what is deemed acceptable use and unacceptable use. An organisation will have an email policy in place aimed at equipping employees with the necessary tools to protect against email-related threats such as phishing attacks. A corporate email policy may also include language covering personal usage of corporate communications systems, informing whether personal emails are accepted.

*Behaviours related to the behavioural set*

Users are susceptible to phishing attacks for a number of reasons. Users may lack the awareness and skills needed to detect phishing attempts and determine between genuine and phishing weblinks, and may not understand security indicators in web browsers (Sheng et al., 2010). By examining factors such as personality traits, user awareness, education, motivation and perception of risk, researchers have been able to form theories of user behaviour when it comes to processing and dismissing or falling victim to phishing attempts.

It is suggested that certain cognitive impulsivity and personality traits affect behavioural responses to genuine and phishing emails (Pattinson et al., 2012). In particular, user extraversion, trust and submissiveness all represent variables that limit the self-efficacy of a user's threat avoidance from phishing emails (Pattinson et al., 2012; Alseadoon et al., 2015). When it comes to executing demands within phishing emails, the behavioural trait 'susceptibility' is a variable that plays an important role in increasing the tendency of a user falling victim to the phishing request (Alseadoon, Othman & Chan, 2015). Studies show that helpfulness, trust and risk behaviours have significant impact on actual email behaviour (Flores et al., 2014).

Awareness and conscientious traits also affect how likely the user is to display target email behaviour. Conscientious users are likely to apply the relevant knowledge to avoid either opening or committing any instructed action as a result of a phishing

email. In general, within simulations, informed users have managed emails better than non-informed users (of course this raises the question of who we understand to be an informed user). Similarly, the more familiar individuals are with computers, the better they managed phishing emails, and in particular, employees with more email experience tended to have more suspicion of phishing emails (Flores et al., 2014.)

A number of additional demographic traits have been shown to affect a user's susceptibility to phishing attacks. Gender and age are two key demographics that affect phishing susceptibility. Although results on gender show mixed results (Kumaraguru et al., 2009), research shows women are more likely than men to be susceptible to phishing; research studies have demonstrated that women click on links more often than men, and are also more likely to enter personal details into input fields. Kumaraguru et al. (2009) note that this appears to be due to the fact that male participants tended to have more technical training and technical knowledge compared to their female counterparts. These findings contrast with findings from another study which suggest that women tend to be less susceptible to a generic attack than men (Flores et al., 2014). This discrepancy suggests that further research into how different demographics affect security behaviour in general and email behaviour in particular is needed. This should include research to refine how gender may affect susceptibility to phishing attacks. Some studies have also found that younger users may be more susceptible to phishing attacks as research participants between the ages of 18-25 were more susceptible compared with other age groups (Sheng et al., 2010; Kumaraguru et al, 2009). Sheng et al. (2009) further suggest that younger participants may be more susceptible due to less experience with internet related activities, less exposure to training materials, and less of an aversion to risks.

While there are a number of gaps in the currently literature (Pattinson et al., 2012), highlighting currently known traits that affect user's susceptibility to phishing scams will enable attempts to target to change specific behavioural attributes.

*Motivating factors driving 'good' behaviour*
Studies have suggested that an individual's threat perception, effectiveness, self-efficacy, perceived severity of threats and perceived susceptibility can positively impact threat avoidance behaviour, whereas safeguard cost can have a negative impact (Arachchilage et al., 2016).

Victims can be manipulated through social engineering techniques which encourage them to act on the instructions given in such emails, however research also shows that an individual's habitual trust and risk behaviour significantly affects actual behaviour during a simulated phishing experiment.

Security awareness methods that target user motivation can enhance a user's avoidance behaviour. In particular, game-based designs have the ability to encourage users to check website URLs as an indicator of website legitimacy. By providing motivation to protect against threats, game-based education delivery has been shown to engage users and lead to an improvement in threat avoidance in post-game tests (Arachchilage et al., 2016).

There is debate as to whether companies should phish their own employees to raise awareness. As 'phishing as a service' is a widely used method in highlighting user susceptibility, some studies suggest that phishing your own employees results in a number of unintended consequences. Research has highlighted that when debriefing participants after a simulated phishing experiment, the majority of employees feel shocked, surprised and even angered at themselves for failing to recognise phishing emails. Many also felt disappointment in their own efficacy (Caputo et al., 2014). Reprimanding employees for clicking on such links might therefore harm the employee-employer trust relationship, reduce productivity and alienate employees, some of whom will be angry at the employer for deceiving them (Caputo et al.,2014). These factors can all play a part in reducing security overall (Renaud and Goucher, 2012; Adams and Sasse, 1999), and may lead to cases where users no longer feel confident to report phishing emails for fear of negative consequences.

*Barriers to behavioural change*

Ineffective phishing awareness delivery has been shown to fail to achieve behavioural change, potentially by failing to challenge employee misconceptions. Research has highlighted cases where almost all employees consider annual company training ineffective, and failed to remember phishing-specific training. Employees claimed this training covered material they already knew and instructed them to act in ways they already did (Caputo et al., 2014). In part, poor training may also be due to the assumption that users are keen to avoid risk, which does not necessarily hold across all circumstances. Kirlappos and Sasse (2012) outline the 'need and greed' principle, highlighting that users can be tempted to click on links in search of a good deal; making them more vulnerable to scams that offer deals that are 'too good to be true'. By not considering the drivers and user motivations, security awareness training offers little protection to this subset of individuals. This research showed much of the advice given through user training was ignored as the indicators were unknown and untrusted by users. This exemplifies that a number of security awareness campaigns may be missing the opportunity to encourage change as they fail to address user misconceptions on scam websites. It has also been pointed out that while security awareness can reduce instances of users clicking on phishing email links, they may also reduce user tendency to click on legitimate links, showing that users are still failing to differentiate between phishing and genuine emails (Sheng et al., 2010).

Employees may also have a false sense of confidence in the company to prevent malicious emails from reading them. Participants in different studies have reported acting differently in the workplace than at home, with the knowledge of dedicated security colleagues, corporate firewalls and cyber security tools such as antivirus software leading to a belief in greater inherent security (Caputo e al., 2014). Caputo et al. (2014) found that these controls made participants more likely to click on links in emails on company computers than on their own personal computers, as they felt protected by the corporate firewall.

Finally, increasingly sophisticated information targeting will make distinguishing genuine emails from phishing emails a more difficult task for potential victims. Ferguson (2015) highlights the issue through the example of a simulation where 80% of participants in a phishing simulation were students expecting a genuine email on their grades. Candidate comments suggested that they would open almost any email relating to grades, stating "any email with the word 'grades' in it gets my immediate attention and action!". The use of social engineering techniques to target particular types of victim suggests an obvious vulnerability.

*Recommendations and 'what works'*

Research has shown that awareness of phishing threats is often not sufficient to change employee behaviour; as an example, a user having attended a training course that covered phishing did not make them less likely to open a phishing email (Ferguson, 2005). While educational and awareness activities pertaining to email environments are of 'utmost importance', IT management must know and identify exactly where to direct and focus these awareness training efforts, according to Steyn et al. (2007). When attempting to determine why some employees appear less susceptible to phishing attempts, role-playing experiments suggest that participants informed about phishing and its risks manage suspicious emails better than those who have not been informed (Pattinson et al., 2012). However, Ferguson (2005) highlights that awareness is necessary but not sufficient as a driver for behavioural change. Broad reviews and analyses of global information security awareness campaigns suggest that as well as awareness, employees must experience both understanding and motivation (Bada and Sasse, 2014), and should have misconceptions challenged and explained (Kirlappos and Sasse, 2012). This highlights a challenge and recommendation that behavioural change should be at the habitual level. Further research supports the hypothesis that conceptual and procedural knowledge positively impacts an employee's self-efficacy when it comes to detecting phishing attempts, enhancing their threat avoidance behaviour (Arachlidge and Love, 2014). The literature therefore supports a recommendation for well-designed end-user security education that informs and enhances employee understanding of phishing and similar attack vectors, and provides a motivation to avoid suspected phishing attempts. As trust and risk behaviours have been shown to affect actual behaviours, it is suggested that training initiatives to develop these behaviours within their security awareness programmes are critical (Flores et al., 2014).

For suggestions on HR implementation in particular, an academic review of existing training models, including the NIST training models, concluded that it is also the method of delivery of training that secures effectiveness, beyond the knowledge conveyed. Needs analysis training is recommended to deliver training where objectives should specify the desired changes in the employees being changed (Brummel et al., 2016). By both classifying employees by role and performing competency modelling, one can determine what each role needs to know and train employees accordingly. This allows employers to identify training gaps, for example whether staff can currently identify phishing attempts, and perform a training evaluation to measure if these gaps have been addressed. A review of effective cyber training approaches suggests that all employees who use computer networks should be trained on a needs based analysis (Beyer and Brummel, 2015). A follow-up exercise should be included within training models to reinforce taught content (Ferguson, 2015).

Companies may also reduce the number of phishing incidents through being mindful of exposing employees to targeted phishing. As research has shown the degree of target information used in a phishing attack increases the likelihood of victims being successfully deceived, it is suggested that organisations consider the benefits of publicly accessible employee email addresses and role titles against the risks of attackers using this information to design more effective phishing messages. (Flores et al., 2014).

**Password behaviour**

This fourth and final behavioural set identified in the literature relates to password behaviour within organisations.

*Definitions and context based on the literature*

Password behaviours in the context of this review will include an array of practices by users of systems, and those who create password policy, such as password composition, security practices and attitudes in relation to passwords (Bryant and Campbell, 2006). A focus on password behaviours is important because although strong authentication techniques are available, corporations continue to use a password-based system to control system access (Tam et al., 2010). Such text-based passwords are often seen as being less secure, easier to predict or guess for an adversary, thereby making it possible for the adversary to impersonate a legitimate user and misuse his or her authority (Shay et al., 2010).

Even the most sophisticated security systems become useless if users mismanage their passwords, or if password policies are not tailored correctly (Furnell et al., 2006). There are many different ways password security can be compromised by adversaries, some of which are unsophisticated and require little knowledge of technology, while others may require high-level technological expertise (Furnell et al., 2006).

Similar to email behaviours, password policy aims to shape password behaviours amongst employees. In general, these are a set of rules established to enhance technological security by ensuring, or at least encouraging individuals to use what is determined to be strong passwords and to employ them in appropriate ways. Such policies may include a requirement for password lengths of at least eight characters, passwords with mixed case/symbols, and the requirement to change passwords regularly (Furnell et al., 2006). It should be noted that these common policies, in reality, are not necessarily good practice (Tam et al., 2010).

In general, users demonstrate knowledge of what forms a strong password (i.e. hard to guess, using a variety of letters, numbers and symbols), as well as inappropriate password practices, such as easily guessed, or commonly used default passwords (Schneier, 2006).

*Behaviours related to the behavioural set*
As previously mentioned, specific behaviours that relate to password behaviour include practices by users of systems, and those who create password policy, such as password composition, security practices and attitudes in relation to passwords (Bryant and Campbell, 2006). Therefore, lots of behaviours and attitudes may affect a user's password behaviour. Tam et al. (2010) specifically approached five password management behaviours: choosing a password for the first time; changing a password; letting someone else use your password; taping passwords next to the computer; and sharing passwords with family, friends or co-workers. 'Choosing a password for the first time' and 'changing a password' are neutral behaviours, in that the actions are not inherently good or bad. It is how the user chooses and updates the password that determines whether the behaviour will have positive or negative effects. The subsequent three, 'letting someone else use your password', 'taping passwords next to the computer', and 'sharing passwords with family, friends or co-workers' are behaviours organisations would want users to avoid. These all represent negative behaviours that are seen as mistakes, and represent instances of poor password management. Research has aimed to understand what motivates each of these behaviours, such as what factors may encourage an employee to deliberately choose week passwords, in order to understand and encourage stronger security behaviour.

There is also the general behavioural theme of password system misuse. Users may be motivated to engage in poor password management. One large-scale study of over half a million users on web-based password behavioural habits demonstrated that there is a high degree of quality passwords and mismanagement (Florencio and Herley, 2007). This may be for a number of reasons, as individuals do, in the end, have a choice about whether they comply or not with password policy (Weirich and Sasse, 2001). For example, users may not comply with password policy because they may not perceive negative consequences within their convenience-security trade-off calculations (Tam et al., 2010).

On the other hand, another important factor to consider when looking at password behavioural themes is misguided password policy. For example, an often-referenced factor is that the challenge of memorising randomised and temporary passwords is both difficult and inconvenient for employees, reducing productivity and prompting workarounds, such as writing passwords down or choosing simpler passwords altogether (Tari and Holden, 2006). Furthermore, researchers have pointed out, that employees have to spend a lot of mental and physical time on some password policies and that this adds to their normal workload (Beautement et al., 2009).

*Motivating factors driving 'good' behaviour*
It is important to address the view that employees are not motivated to behave in a secure manner. In their study, Adams and Sasse (1999) found that the majority of users were in fact security conscious, as long as the users felt the need for such behaviours (for example owing to external threats). However, it should be noted that in the literature, there seems to be a pattern by some researchers suggesting that users intentionally or through negligence are a great threat for information security (Safa et al., 2015).

The literature suggests that users understand the difference between good and bad behaviour, and that motives behind password selection and password management are complex and significant in shaping behaviours. Some users are more motivated by privacy issues rather than security. Users are also motivated by security and convenience simultaneously, and will make a trade-off between them when determining a password. This trade-off often determines password quality meaning that users will choose a strong password only if they are willing to sacrifice convenience; awareness is not sufficient (Tam et al., 2010).

*Barriers to behavioural change*

Research shows that awareness of password policy and the consequences of passwords being compromised is not always enough to drive 'good' password behaviour. Studies have shown that users may also be aware of what makes a good or bad password, but still not be motivated to comply with this (Tam et al., 2010). The Tam et al. (2010) study also found that this lack of motivation was because users did not see any immediate negative consequences of engaging in 'bad' security behaviour, or because of a security-convenience trade-off. This is therefore a barrier to behavioural change, as clearly employees need a higher level of motivation, rather than just awareness.

As previously noted, password policies can often be too demanding for employees to be able to cope with, or may interfere with employee work productivity. This therefore is a barrier to behavioural change. Research shows that in most cases, those who use workarounds to circumvent password policies are not 'black hat' hackers, but employees trying to do their jobs efficiently despite certain policies (Koppel et al., 2015).

Furthermore, studies have shown that in order to deal with the demand of keeping track of many passwords for different accounts, users write down and reuse passwords, with there being diverse individual behaviour surrounding this (Stobert and Biddle, 2014). It has been further suggested by researchers, that this 'bad' password behaviour is actually a rational response to the high demands of varying policies (Stobert and Biddle, 2014). These findings are corroborated by many different studies. For example, Inglesant and Sasse (2010) found that employees were concerned about security, but that policies were inflexible and did not match their capabilities. Such studies demonstrate that these password policies can place demands on users which impact negatively on their productivity.

*Recommendations and 'what works'*

Password policies must strike the right balance between maximising the security of the organisation, while minimising user frustration and maximising user usability (Komanduri et al., 2011). Password policies will continually have to be adapted and must be responsive to user behavioural patterns, the emergence of new threats, and will be dependent on the nature of the systems for which passwords are used.

(a) <u>Direct behavioural change in terms of encouraging behaviour change.</u>

Studies have found that for sensitive applications such as online banking log-ins, the concept of time-frames (i.e. whether the password will change immediately or in the future) will affect an individual's perception of the security trade-offs, where weaker passwords were chosen for immediate change, and stronger passwords chosen for those due to change in the future (Tam et al., 2010).

(b) <u>Indirect behavioural change through design mechanisms</u>

Surveys have revealed that user-accessibility and communication both discourage strong password behaviours, emphasising the responsibility on designers of security mechanisms to encourage responsible password behaviour by design. Responses suggested reducing the effort involved and highlighting where and why there is a need for a strong password within a security design, users are better motivated to create stronger passwords (Adams abd Sasse, 1999). Angela Sasse's ongoing research on password use supports this conclusion, while related research suggests involving user collaboration in participatory research leads to the desirable outcome of both a better understanding of and relation to an organisation's security requirements (Kani-Zabihi and Helmhout, 2011).

The National Cyber Security Centre (2016) has also previously released recommendations for password policy to improve password behaviour, and is about to release an updated version. Recommendations revolve around dramatic simplification of complicated approaches to password policy, as highlighted at the beginning of this theme.

# SECTION FOUR
# SUMMARISING 'WHAT WORKS'

This section provides a brief summary of the 'what works' sections from each of the four behavioural sets. To preface this, it is notable that, currently, there is not enough research or empirical data, and there are too many confounding variables from organisation to organisation to conclude that certain behavioural interventions will work in every organisational setting.

Awareness campaigns have been mentioned in almost every 'what works' section in this report (Rhee et al., 2009). Awareness campaigns have been shown to be effective for education and training (Cone et al., 2007). However, some studies have shown that even if these awareness campaigns are effective in boosting knowledge in cyber security, this does not necessarily mean the awareness campaigns actually have an impact on user behaviours (Albrechtsen, 2007). Researchers have shown that campaigns might be more effective if they involve more motivation, or take a user-involvement approach. Bada and Sasse (2014) note that efforts to change security behaviour necessitate a lot more than just giving information about risks and correct behaviours. It is argued that employees must not only be able to understand and apply the advice they are given, but they must be willing to do apply such advice, which requires attitudinal change (Nurse, 2015).

Techniques for attitudinal change include persuasion techniques such as fear appeals. Fear appeals should aim to address aspects of attitudes such as employee optimism bias when it comes to some aspects of security (Rhee et al., 2005). Again, there is also research that counters this and suggests that employee knowledge of security, for example phishing, does enhance user behaviour (Arachchilage and Love, 2014; Alqahtani, 2017). Overall therefore, research into awareness campaigns is relatively mixed and, as previously mentioned, security awareness plans are limited to organisations that do not already have a good working knowledge of cyber security.

The use of sanctions and rewards in order to encourage behavioural change also comes up in a few of the themes. However, the use of both is a contested topic in the research and literature. Cheng et al. (2013) recommend increased severity of sanctions, while Pahnil et al. (2007) conclude that sanctions have no significant impact on intention to comply. Kirlappos et al. (2014) instead suggest that sanctions are not an effective answer, and signpost to policy design. While this may suggest a shift in emphasis to focus on positive reinforcement, rewards have been shown to result in inferior information security decision-making (Parsons et al., 2015). In this case, the authors suggest it may be that different types of employees respond positively to different approaches, suggesting inappropriate use of rewards would be detrimental to compliance. These remain open questions and raise issues for further research in this area.

As highlighted in the security culture section of this review, the use of top management to support cyber security behaviours and to improve cyber security in general, has been found to be effective (Werlinger et al., 2009). This finding can easily be applied to all four behavioural sets. Researchers have suggested that security vulnerability, and risk analysis reports can be used to convince top management about the importance of cyber security, and want to lead on this front (Werlinger et al., 2009).

**Limitations of the literature**

Overall, in relation to the four 'what works' sections of this report, it is clear that more research is needed to assess the reliability and validity of all the behavioural interventions mentioned. Furthermore, longitudinal studies and more studies to test the reliability of different interventional methods over time, and whether they have lasting effects on everyday cyber security in organisations. Additionally, there is a need to establish metrics in order for researchers to be able to consistently measure the effects behavioural interventions on behaviour. Unlike in medicine or clinical psychology, where randomised controlled trials are the gold standard, behavioural interventions in organisations will prove more difficult when trying to establish clear cut results.

Additionally, in certain areas of this review, the research findings are mixed, leading us to have significant reservations about recommending some 'solutions' or approaches to change security behaviours. For example, while a number of studies have examined reward and punishment, and the effectiveness of sanctions in particular, conclusions and prescriptions vary significantly (Cheng et al., 2013; Pahnila et al., 2007). Similarly, the use of awareness campaigns is also disputed (Bada and Sasse, 2014).

Particularly when reviewing literature on compliance behaviours, a significant number of papers used intention to comply as the dependent variable, raising the question of whether intention indicates actual behaviour (acknowledged by Cheng et al., 2013). While generally there is support that intention may be used as a predictor of actual behaviour, there is no guarantee that surveyed individuals would behave as they have indicated. This forms part of the wider challenge when it comes to measuring compliance; not every type of employee violation can be evidenced, and this represents a major challenge beyond academia and for security employees. This difficulty in exploring and measuring actual behaviour can be seen in the field of research on phishing behaviour. Little empirical research has taken place through real phishing experiments in part due to the ethical questions of launching a phishing campaign on employees without providing a proper debrief. More specifically, very little research has focused on individuals in a workplace setting, which may be due to the difficulty involved in persuading organisational managers to participate in studies in which their employees' performances are being tested. (Flores et al., 2014). Where research on factors affecting susceptibility to phishing attacks have taken place, their applicability is limited by the fact they large rely on the data of university students, with small sample sizes limiting possible statistical analysis on possible determinants of susceptibility (Flores et al., 2014).

Finally, and importantly, the lack of theoretical underpinnings and critical reflection and engagement with the topic of security behaviours in most of the studies, makes the evidence base inconsistent and poor. As a result, it is not possible to recommend any conclusive suggestions as to 'what works' as we would be basing such recommendations on some pretty disconnected, isolated and more or less informative pieces of literature and studies.

**Directions for future research**

In general, the overarching direction for research is the need for more behavioural change research in this area, and a deeper engagement with the fundamental principles of security. This research needs to repeat existing research in different environments to determine the validity and reliability of existing methods to a greater extent. Furthermore, this future research should aim to use psychological, sociological, and economic theory to aid, add to, or create new, behavioural interventions. It has been noted in the research that these disciplines, such as social psychology, have thus far been underused in research into cyber security (Thackray et al., 2016), and that there are a lot of related theories that still need to be examined for their applications to this area.

More specifically, incoming regulations pose a significant challenge to information security management within organisations. New requirements, such as the GDPR (effective from May 2018) have prompted major change programmes across industry. The effects of the regulation on security behaviours in the workplace have not been specifically examined. As each relevant aspect of the regulation is integrated into the workplace and organisational policies, there are significant research opportunities to study the effect of GDPR on security behaviours.

There is a need for more research on behavioural differences between types of employee, or within different organisational environments which may display different behaviours towards cyber security issues. There is research in the existing literature on gender, age, and income differences (Akman and Mishra, 2010), however, the findings are inconclusive and other factors are in need of investigation. For example, while research has highlighted the relationship between job satisfaction and employee intention to comply with security policies and procedures, this relationship is contingent on position, tenure and industry. The relationship is stronger for those employees with non-technical positions, in non-IT industries and for those with less tenure in their organisations. Future research should attempt to work out the causes of differences between job satisfaction, believes and attitudes between employees in different environments. Likewise, as mentioned within the above review of 'inter-group dynamics' research, there is significant scope to expand on cases where conflict may actually strengthen team performance and lead to greater security overall. There is plenty of opportunity to explore personality types and how various types of employee differ in their approaches to, for example, compliance or phishing (Pattinson et al, 2012). With more research and empirical evidence will come clearer methodologies on increasing employee intentions to comply, whether relating to this specific case of job satisfaction, inter-group dynamics or across a number of other factors.

The need for effective security education training was highlighted across the literature. Several academics highlighted opportunities for needs-based analysis to be incorporated into educational policy. Game-based educational delivery showed promise in studies using prototypes, teaching users how to recognise phishing emails by examining website URLs. There are opportunities to explore how games can be employed to teach other key security lessons, encouraging the user to engage to protect themselves (Arachilage et al., 2015).

More broadly, a common theme within the academic literature was a need to continue to bring together diverse approaches to information security behaviours. The theories used in the research of information security draw on a number of distinct fields with evolving theories. The ongoing development and innovation within this space relies on engagement with emerging theories across psychology, criminology, management and information security research areas, and potentially developing new theories specifically relevant to cyber security behaviours.

# REFERENCES

Adams, A. and Blandford, A., 2005. Bridging the gap between organizational and user perspectives of security in the clinical domain. *International Journal of Human-Computer Studies*, 63(1), pp.175-202.

Adams, A. and Sasse, M.A., 1999. Users are not the enemy. *Communications of the ACM*, *42*(12), pp.40-46.

Akman, I. and Mishra, A., 2010. Gender, age and income differences in internet usage among employees in organizations. *Computers in Human Behavior*, *26*(3), pp.482-490.

Albrechtsen, E., 2007. A qualitative study of users' view on information security. *Computers & Security, 26*(4), pp.276-289.

Albrechtsen, E. and Hovden, J., 2009. The information security digital divide between information security managers and users. *Computers & Security*, 28(6), pp.476-490.

Allport, G.W., 1954. The nature of prejudice.

Alseadoon, I., Othman, M.F.I. and Chan, T., 2015. What Is the Influence of Users' Characteristics on Their Ability to Detect Phishing Emails?. In Advanced Computer and Communication Engineering Technology (pp. 949-962). Springer, Cham.

Alqahtani, F.H., 2017. Developing an Information Security Policy: A Case Study Approach. *Procedia Computer Science, 124*, pp.691-697. Arachchilage, N.A.G. and Love, S., 2014. Security awareness of computer users: A phishing threat avoidance perspective. *Computers in Human Behavior*, 38, pp.304-312.

Arachchilage, N.A.G., Love, S. and Beznosov, K., 2016. Phishing threat avoidance behaviour: An empirical investigation. *Computers in Human Behavior*, 60, pp.185-197.

E. Aronson, D.W. Timothy, R.M. Akert, Social Psychology, Prentice Hall, Upper Saddle River, NJ, 2010 p. 2010.

Bada, M. and Sasse, A., 2014. Cyber Security Awareness Campaigns: Why do they fail to change behaviour?.

Beautement, A., Sasse, M.A. and Wonham, M., 2009, August. The compliance budget: managing security behaviour in organisations. In *Proceedings of the 2008 New Security Paradigms Workshop* (pp. 47-58). ACM.

Becker, I., Parkin, S. and Sasse, M.A., 2017. Finding Security Champions in Blends of Organisational Culture. *Proc. USEC*, pp.1-11.

Bélanger, F., Collignon, S., Enget, K. and Negangard, E., 2017. Determinants of early conformance with information security policies. Information & Management.

Benson, V., McAlaney, J. and Frumkin, L.A., 2018. Emerging Threats for the Human Element and Countermeasures in Current Cyber Security Landscape. *In Psychological and Behavioral Examinations in Cyber Security* (pp. 266-271). IGI Global.

Beyer, R.E., Integritas, L.L.C. and Brummel, B.J., 2015. Implementing effective cyber security training for end users of computer networks. SHRM-SIOP Science of HR Series: Promoting Evidence-Based HR.

Bradley, B.H., Anderson, H.J., Baur, J.E. and Klotz, A.C., 2015. When conflict helps: Integrating evidence for beneficial conflict in groups and teams under three perspectives. *Group Dynamics: Theory, Research, and Practice, 19*(4), p.243.

Brown, R., 1988. *Group processes: Dynamics within and between groups.* Basil Blackwell.

Bryant, K. and Campbell, J., 2006. User behaviours associated with password security and management. *Australasian Journal of Information Systems*, 14(1).

Bulgurcu, B., Cavusoglu, H. and Benbasat, I., 2010. Information security policy compliance: an empirical study of rationality-based beliefs and information security awareness. *MIS quarterly*, 34(3), pp.523-548.

Caputo, D.D., Pfleeger, S.L., Freeman, J.D. and Johnson, M.E., 2014. Going spear phishing: Exploring embedded training and awareness. IEEE Security & Privacy, 12(1), pp.28-38.

Cheng, L., Li, Y., Li, W., Holm, E. and Zhai, Q., 2013. Understanding the violation of IS security policy in organizations: An integrated model based on social control and deterrence theory. *Computers & Security*, 39, pp.447-459.

Chen, C.C., Dawn Medlin, B. and Shaw, R.S., 2008. A cross-cultural investigation of situational information security awareness programs. *Information Management & Computer Security,* 16(4), pp.360-376.

Chen, Y., Ramamurthy, K. and Wen, K.W., 2012. Organizations' information security policy compliance:

Stick or carrot approach?. J*ournal of Management Information Systems,* 29(3), pp.157-188.

Cone, B.D., Irvine, C.E., Thompson, M.F. and Nguyen, T.D., 2007. A video game for cyber security training and awareness. *Computers & Security, 26*(1), pp.63-72.

Coventry, L., Briggs, P., Jeske, D. and van Moorsel, A., 2014, June. Scene: A structured means for creating and evaluating behavioral nudges in a cyber security environment. In *International Conference of Design, User Experience, and Usability* (pp. 229-239). Springer, Cham.

D'Arcy, J. and Greene, G. 2014. Security culture and the employment relationship as drivers of employees' security compliance. *Information Management and Computer Security*, 22(5), pp.474-489.

De Guinea, A.O. and Markus, M.L., 2009. Why break the habit of a lifetime? Rethinking the roles of intention, habit, and emotion in continuing information technology use. *MIS Quarterly*, pp.433-444.

Dhamija, R. and Tygar, J.D., 2005, July. The battle against phishing: Dynamic security skins. In Proceedings of the 2005 symposium on Usable privacy and security (pp. 77-88). ACM.

Farahmand, F., Atallah, M.J. and Spafford, E.H., 2013. Incentive alignment and risk perception: An information security application. *IEEE Transactions on Engineering Management*, *60*(2), pp.238-246.

Ferguson, A.J., 2005. Fostering e-mail security awareness: The West Point carronade. Educase Quarterly, 28(1), pp.54-57.

Florencio, D. and Herley, C., 2007, May. A large-scale study of web password habits. In *Proceedings of the 16th international conference on World Wide Web* (pp. 657-666). ACM.

Floyd, D.L., Prentice-Dunn, S. and Rogers, R.W., 2000. A meta-analysis of research on protection motivation theory. *Journal of applied social psychology*, *30*(2), pp.407-429.

Fritsch, C., 2015. Data Processing in Employment Relations; Impacts of the European General Data Protection Regulation Focusing on the Data Protection Officer at the Worksite. In *Reforming European Data Protection Law* (pp. 147-167). Springer, Dordrecht.

Furnell, S.M., Jusoh, A. and Katsabas, D., 2006. The challenges of understanding and using security: A survey of end-users. *Computers & Security*, 25(1), pp.27-35.

Gabriel, T. and Furnell, S., 2011. Selecting security champions. *Computer Fraud & Security*, 2011(8), pp.8-12.

Gaertner, S.L., Dovidio, J.F., Banker, B.S., Houlette, M., Johnson, K.M. and McGlynn, E.A., 2000. Reducing intergroup conflict: From superordinate goals to decategorization, recategorization, and mutual differentiation. *Group Dynamics: Theory, Research, and Practice*, *4*(1), p.98.

Gcaza, N. and von Solms, R., 2017, May. Cybersecurity Culture: An Ill-Defined Problem. In *IFIP World Conference on Information Security Education* (pp. 98-109). Springer, Cham.

Glaspie, H.W. and Karwowski, W., 2017, July. Human Factors in Information Security Culture: A Literature Review. In *International Conference on Applied Human Factors and Ergonomics* (pp. 269-280). Springer, Cham.

Grugulis, I. and Vincent, S., 2009. Whose skill is it anyway? 'soft'skills and polarization. *Work, employment and society*, *23*(4), pp.597-615.

Guo, K.H., 2013. Security-related behavior in using information systems in the workplace: A review and synthesis. *Computers and Security*, 32(1), pp.242-251.

Guynes, C.S. and Windsor, J., 2012. Security awareness programs. *The Review of Business Information Systems (Online)*, 16(4), p.165.

Han, J., Kim, Y.J. and Kim, H., 2017. An integrative model of information security policy compliance with psychological contract: Examining a bilateral perspective. Computers & Security, 66, pp.52-65.

Harris, L.C. and Ogbonna, E., 1998. Employee responses to culture change efforts. *Human Resource Management Journal*, *8*(2), pp.78-92.

S. Hazari, W. Hargrave, B. Clenney, An empirical investigation of factors influenc- ing information security behavior, Journal of Information Privacy and Security 4 (4), 2009, pp. 3–20.

Helmreich, R.L. and Foushee, H.C., 1993. *Why crew resource management? Empirical and theoretical bases of human factors training in aviation*. Academic Press.

Herath, T. and Rao, H.R., 2009. Encouraging information security behaviors in organizations: Role of penalties, pressures and perceived effectiveness. *Decision Support Systems*, *47*(2), pp.154-165.

HM Government and PWC. (2015). INFORMATION SECURITY BREACHES SURVEY 2015. London: HM Government, pp.1-8.

Hogg, M.A., Van Knippenberg, D. and Rast, D.E., 2012. Intergroup leadership in organizations: Leading across group and organizational boundaries. *Academy of Management Review*, 37(2), pp.232-255.

Hovav, A. and D'Arcy, J., 2012. Applying an extended model of deterrence across cultures: An investigation of information systems misuse in the US and South Korea. *Information & Management,* 49(2), pp.99-110.

Hsu, J.S.C., Shih, S.P., Hung, Y.W. and Lowry, P.B., 2015. The role of extra-role behaviors and social controls in information security policy effectiveness. *Information Systems Research*, 26(2), pp.282-300.

Hu, Q., Dinev, T., Hart, P. and Cooke, D., 2012. Managing employee compliance with information security policies: The critical role of top management and organizational culture. *Decision Sciences*, *43*(4), pp.615-660.

Ifinedo, P., 2014. Information systems security policy compliance: An empirical study of the effects of socialisation, influence, and cognition. *Information & Management*, 51(1), pp.69-79.

Inglesant, P.G. and Sasse, M.A., 2010, April. The true cost of unusable password policies: password use in the wild. In *Proceedings of the SIGCHI Conference on Human Factors in Computing Systems* (pp. 383-392). ACM.

Karyda, M., 2017. Fostering Information Security Culture In Organizations: A Research Agenda.

Kani-Zabihi, E. and Helmhout, M., 2011, October. Increasing service users' privacy awareness by introducing on-line interactive privacy features. In *Nordic Conference on Secure IT Systems* (pp. 131-148). Springer, Berlin, Heidelberg.

Kirlappos, I. and Sasse, M.A., 2012. Security education against phishing: A modest proposal for a major rethink. IEEE Security & Privacy, 10(2), pp.24-32.

Kirlappos, I., Parkin, S. and Sasse, M.A., 2014. Learning from "Shadow Security": Why understanding non-compliance provides the basis for effective security.

Kolkowska, E. 2011. Security subcultures in an organization-exploring value conflicts. *European Conference on Information Systems,* ECIS 2011 Proceedings, (p. 237).

Komanduri, S., Shay, R., Kelley, P.G., Mazurek, M.L., Bauer, L., Christin, N., Cranor, L.F. and Egelman, S., 2011, May. Of passwords and people: measuring the effect of password-composition policies. In *Proceedings of the SIGCHI Conference on Human Factors in Computing Systems* (pp. 2595-2604). ACM.

Koppel, R., Smith, S.W., Blythe, J. and Kothari, V., 2015. Workarounds to computer access in healthcare organizations: you want my password or a dead patient?. In *ITCH* (pp. 215-220).

Krombholz, K., Hobel, H., Huber, M. and Weippl, E., 2015. Advanced social engineering attacks. *Journal of Information Security and applications*, 22, pp.113-122.

Kumaraguru, P. Cranshaw, J., Acquisti, A., Cranor, L., Hong, J., Blair, M. A., and Pham, T. School of Phish: A Real-World Evaluation of Anti-Phishing Training. In the Proceedings On Usable Privacy and Security, 2009.

Martin, J., Frost, P.J. and O'Neill, O.A. 2006. *"Organizational culture: beyond struggles for intellectual dominance",* in Clegg, S., Hardy, C., Nord, W. and Lawrence, T. (Eds), Handbook of Organization Studies, 2nd ed., Sage Publications, London, pp. 725-753.

Martins, C., Oliveira, T. and Popovič, A., 2014. Understanding the Internet banking adoption: A unified theory of acceptance and use of technology and perceived risk application. *International Journal of Information Management*, 34(1), pp.1-13.

McSweeney, B., 1999. *Security, identity and interests: a sociology of international relations* (Vol. 69). Cambridge University Press.

Merete Hagen, J., Albrechtsen, E. and Hovden, J., 2008. Implementation and effectiveness of organizational information security measures. *Information Management & Computer Security,* 16(4), pp.377-397.

Mitnick, K.D., Simon, W.L. and Wozniak, S., 2006. The Art of Deception: Controlling the Human Element of Security. 2002. Paperback ISBN 0-471-23712-4.

Nandi, A.K., Medal, H.R. and Vadlamani, S., 2016. Interdicting attack graphs to protect organizations from cyber attacks: A bi-level defender–attacker model. *Computers and Operations Research*, 75, pp.118-131.

National Cyber Security Centre. (2016). *Password Guidance: Simplifying Your Approach - NCSC Site.* [online] Available at: https://www.ncsc.gov.uk/guidance/password-guidance-simplifying-your-approach [Accessed 17 Apr. 2018]

Nelson, R.E., 1989. The strength of strong ties: Social networks and intergroup conflict in organizations. *Academy of Management Journal*, *32*(2), pp.377-401.

Nurse, J., 2015. Cyber Security Awareness Campaigns: Why do they fail to change behaviour?.

Pahnila, S., Siponen, M. and Mahmood, A., 2007, January. Employees' behavior towards IS security policy compliance. In *System sciences, 2007. HICSS 2007. 40Th annual hawaii international conference on* (pp. 156b-156b). IEEE.

Parkin, S.E., van Moorsel, A. and Coles, R., 2009, October. An information security ontology incorporating human-behavioural implications. *Proceedings of the 2nd International Conference on Security of Information and Networks* (pp. 46-55). ACM.

Parsons, K., McCormac, A., Butavicius, M. and Ferguson, L., 2010. Human factors and information security: individual, culture and security environment. *Australian Defence, Science and Technology Organisation.*

Parsons, K.M., Young, E., Butavicius, M.A., McCormac, A., Pattinson, M.R. and Jerram, C. 2015. The influence of organizational information security culture on information security decision making. *Journal of Cognitive Engineering and Decision Making,* 9(2), pp.117-129.

Pattinson, M., Jerram, C., Parsons, K., McCormac, A. and Butavicius, M., 2012. Why do some people manage phishing e-mails better than others?. Information Management & Computer Security, 20(1), pp.18-28.

Pettigrew, T.F., 1998. Intergroup contact theory. *Annual review of psychology,* 49(1), pp.65-85.

Renaud, K. and Goucher, W., 2012. Health service employees and information security policies: an uneasy partnership? *Information Management and Computer Security*, 20(4), pp.296-311.

Rhee, H.S., Kim, C. and Ryu, Y.U., 2009. Self-efficacy in information security: Its influence on end users' information security practice behavior. *Computers & Security*, *28*(8), pp.816-826.

Rhee, H.S., Ryu, Y. and Kim, C.T., 2005. I am fine but you are not: Optimistic bias and illusion of control on information security. ICIS 2005 Proceedings, p.32.

Rocha Flores, W., Holm, H., Svensson, G., & Ericsson, G. (2014). Using phishing experiments and scenario-based surveys to understand security behaviours in practice. Information Management & Computer Security, 22(4), 393-406.

Ruighaver, A.B., Maynard, S.B. and Chang, S., 2007. Organisational security culture: Extending the end-user perspective. *Computers and Security*, 26(1), pp.56-62.

Safa, N.S., Sookhak, M., Von Solms, R., Furnell, S., Ghani, N.A. and Herawan, T., 2015. Information security conscious care behaviour formation in organizations. *Computers & Security*, *53*, pp.65-78.

Schlienger, T. and Teufel, S., 2003. Information security culture-from analysis to change. *South African Computer Journal*, 2003(31), pp.46-52.

Schneier, B., 2006. Beyond fear: thinking sensibly about security in an uncertain world. New York: Springer Science, Business Media, LLC.

Shay, R., Komanduri, S., Kelley, P.G., Leon, P.G., Mazurek, M.L., Bauer, L., Christin, N. and Cranor, L.F., 2010, July. Encountering stronger password requirements: user attitudes and behaviors. *In Proceedings of the Sixth Symposium on Usable Privacy and Security* (p. 2). ACM.

Sheng, S., Holbrook, M., Kumaraguru, P., Cranor, L.F. and Downs, J., 2010, April. Who falls for phish?: a demographic analysis of phishing susceptibility and effectiveness of interventions. In Proceedings of the SIGCHI Conference on Human Factors in Computing Systems (pp. 373-382). ACM.

Sheng, S., Magnien, B., Kumaraguru, P., Acquisti, A., Cranor, L.F., Hong, J. and Nunge, E., 2007, July. Anti-phishing phil: the design and evaluation of a game that teaches people not to fall for phish. In Proceedings of the 3rd symposium on Usable privacy and security (pp. 88-99). ACM.

Sherif, M., 1958. Superordinate goals in the reduction of intergroup conflict. *American journal of Sociology, 63*(4), pp.349-356.

Siponen, M. and Willison, R., 2009. Information security management standards: Problems and solutions. *Information & Management*, 46(5), pp.267-270.

Sommestad, T., Hallberg, J., Lundholm, K. and Bengtsson, J., 2014. Variables influencing information security policy compliance: a systematic review of quantitative studies. *Information Management and Computer Security,* 22(1), pp.42-75.

Soomro, Z.A., Shah, M.H. and Ahmed, J., 2016. Information security management needs more holistic approach: A literature review. *International Journal of Information Management*, 36(2), pp.215-225.

Stanton, Jeffrey, et al. "Behavioral information security: two end user survey studies of motivation and security practices." AMCIS 2004 Proceedings (2004): 175.

Steyn, T., Kruger, H.A. and Drevin, L., 2007, May. Identity theft—Empirical evidence from a Phishing exercise. In IFIP International Information Security Conference (pp. 193-203). Springer, Boston, MA.

Stobert, E. and Biddle, R., 2014, July. The password life cycle: user behaviour in managing passwords. In *Proc*. SOUPS.

Straub Jr, D.W., 1990. Effective IS security: An empirical study. *Information Systems Research,* 1(3), pp.255-276.

Tam, L., Glassman, M. and Vandenwauver, M., 2010. The psychology of password management: a tradeoff between security and convenience. *Behaviour & Information Technology,* 29(3), pp.233-244.

Tari, F., Ozok, A. and Holden, S.H., 2006, July. A comparison of perceived and real shoulder-surfing risks between alphanumeric and graphical passwords. In Proceedings of the second symposium on Usable privacy and security (pp. 56-66). ACM.

Thackray, H., McAlaney, J., Dogan, H., Taylor, J. and Richardson, C., 2016, July. Social Psychology: An under-used tool in Cybersecurity. In *Proceedings of the 30th International BCS Human Computer Interaction Conference: Companion Volume* (p. 17). BCS Learning & Development Ltd.

Thaler, R.H., & Sunstein, C.R. Nudge. Improving Decisions About Health, Wealth and Happiness. Penguin. (2008).

Theoharidou, M. and Gritazalis, D., 2007. Common body of knowledge for information security. *IEEE Security & Privacy*, 5(2), pp.1540-7993.

Thomson, K. and Van Niekerk, J., 2012. Combating information security apathy by encouraging prosocial organisational behaviour. *Information Management & Computer Security*, 20(1), pp.39-46.

Turland, J., Coventry, L., Jeske, D., Briggs, P. and van Moorsel, A., 2015, July. Nudging towards security: Developing an application for wireless network selection for android phones. In *Proceedings of the 2015 British HCI conference*(pp. 193-201). ACM.

Van Niekerk, J.F. and Von Solms, R., 2010. Information security culture: A management perspective. *Computers & Security*, 29(4), pp.476-486.

Vance, A., Siponen, M. and Pahnila, S., 2012. Motivating IS security compliance: insights from habit and protection motivation theory. *Information & Management*, 49(3-4), pp.190-198.

Von Solms, R. and Von Solms, B., 2004. From policies to culture. *Computers & Security,* 23(4), pp.275-279.

Watson, R.T. and Brancheau, J.C., 1991. Key issues in information systems management: an international perspective. *Information & Management*, 20(3), pp.213-223.

Weirich, D. and Sasse, M.A., 2001, September. Pretty good persuasion: a first step towards effective password security in the real world. In *Proceedings of the 2001 workshop on New security paradigms* (pp. 137-143). ACM.

Werlinger, R., Hawkey, K. and Beznosov, K., 2009. An integrated view of human, organizational, and technological challenges of IT security management. *Information Management & Computer Security*, 17(1), pp.4-19.

West, R., 2008. The psychology of security. *Communications of the ACM*, 51(4), pp.34-40.

Whitty, M., Doodson, J., Creese, S. and Hodges, D., 2015. Individual differences in cyber security behaviors: an examination of who is sharing passwords. *Cyberpsychology, Behavior, and Social Networking,* 18(1), pp.3-7.

31